\documentclass[fleqn,twoside]{article}
\usepackage{espcrc2}

\usepackage{graphicx}
\usepackage[figuresright]{rotating}
\usepackage{epsfig}
\usepackage{url}
\newcommand{\AmS}{{\protect\the\textfont2
  A\kern-.1667em\lower.5ex\hbox{M}\kern-.125emS}}
\title{  TAUOLA, TAUOLA universal interface  PHOTOS and MC-TESTER: Status Report}

\author{ Z. W\c{a}s\address[MCSD]{Institute of Nuclear Physics, Polish Academy of Sciences,\\
         ul. Radzikowskiego 152, 31-342 Cracow, Poland}
        \thanks{  Supported in part  by the Polish Government grant
 N202 06434 (2008-2010) and  EU-RTN Programme, Contract No.
MRTN--CT-2006-035482, \lq\lq Flavianet\rq\rq. }}

\begin{document}

\begin{abstract}
The status of Monte Carlo programs for the simulation of $\tau$-lepton production and decay in 
high-energy accelerator experiments is reviewed. 
No basic changes in the organization of the programs were necessary
since the previous $\tau$-lepton conference in 2006. Minor in size, 
but practical, extensions for the use of the programs in high precision mixed languages
enviromens
are  being introduced:
 (i) For the {\tt TAUOLA} Monte Carlo generator of $\tau$-lepton decays,  
automated and simultaneous use of  many versions of 
form-factors for the calculation of optional weights  for fits was discussed. 
A pilot example of its use was presented.
(ii) New tests for {\tt PHOTOS} Monte Carlo for QED bremsstrahlung 
in  $W$ decays were shown.
(iii) Prototype version of 
the  {\tt TAUOLA universal interface} based on {\tt HepMC} (the {\tt C++} event record)  was mentioned. Its tests with the help of {\tt MC-TESTER} were discussed.

Presented results illustrate  the status of the projects performed in 
collaboration with  Vladimir Cherepanov, Nadia Davidson, Piotr Golonka,
Gizo Nanava, Tomasz Przedzi\'nski   El\.zbieta Richter-W\c{}as and others.

\vspace{3mm}
\centerline{ \it Presented at International workshop on Tau Lepton Physics, TAU08 
Novosibirsk, Russia September, 2008}
\vspace{1pc}
\centerline{preprint \hskip 1 cm  IFJPAN-IV-2008-10}
\vspace{1pc}
\end{abstract}

\maketitle

% Because \thanks uses symbolic footnote marks, we need to reset the counter.
\setcounter{footnote}{0}
 
%%%%%%%%%%%%%
\section{Introduction}

%{\tt /home/wasm/y2004/transparencje/kontrybucja /TAU04-ZWas } 
%%%%%%%%%%%%%%%%%%%%%%%%%%%%%%%%%%%%%%%%%%%%%%%%%%%%%%%%%%%%%%%%%%%%%%%%%%%%%%%%%%%%%%%%%%%%%
The {\tt TAUOLA} package
\cite{Jadach:1990mz,Jezabek:1991qp,Jadach:1993hs,Golonka:2003xt}  for the simulation 
of $\tau$-lepton decays and  
{\tt PHOTOS} \cite{Barberio:1990ms,Barberio:1994qi} for the simulation of radiative corrections
in decays, are computing
projects with a rather long history. Written and maintained by 
well-defined authors, they nonetheless migrated into a wide range
of applications where they became ingredients of 
complicated simulation chains. As a consequence, a large number of
different versions are presently in use.
From the algorithmic point of view, they often
differ only in a few small details, but incorporate many specific results from distinct
$\tau$-lepton measurements or phenomenological projects. 
Such versions were mainly maintained (and will remain so) 
by the experiments taking precision data on $\tau$ leptons. On the other hand,
 many new applications were developed  recently,  often requiring
a program interface to other packages  (e.g. generating events for LHC, LC, 
Belle or BaBar physics processes). The programs structure,
prepared for  the convenience of users,  was presented during previous 
$\tau$ conferences, 
and we will not repeat it here. 

This time, let us concentrate on physics oriented results and special 
techniques being developped and used for fits, over the years.
They may be worth some attention now, when high precision,
high statistic  data are being analyzed. 
{\tt FORTRAN} to {\tt C++}  migration  is not as simple as one may think. 
On one hand it is generally believed that 
large projects are easier to maintain in more modern programming frameworks. 
On the other hand, {\tt FORTRAN } software
can not be abandoned. It usually carries a lot of physics expertise.
It is also impractical  to replace the software environment of mature experiments
with something new.
Fortunately, co-existence of the two languages is not a problem, at least 
not from 
the software point of view.

Our presentation is organized as follows. 
Section 2  is devoted to the discussion  of optional 
weights in {\tt TAUOLA} and their use for fits to the data already at the level of comparison with raw data.
In section 3 we present some new results for the simulation  
with {\tt PHOTOS} 
of radiative corrections in decays. 
Section 4 is devoted to {\tt MC-TESTER}; the program which can be used for 
semi-automatic comparisons of simulation samples originating from
different programs. The prototype, {\tt C++} based,  
{\tt TAUOLA universal interface} is mentioned in that Section too.

Because of the limited space of the contribution, 
and sizable amount of other physically interesting 
results, some of them will be excluded from conference 
proceedings. Let us hope that they will find place in 
future works, possibly with collaborators mentioned in the Abstract.
For these works,  the present paper may serve as an announcement. 

 %Let us mention only references \cite{Richter-Was:2004jf,Chankowski:2004tb}.
%%%%%%%%%%%%%%%%%%%%%%%%%%%%%%%%%%%%%%%%%%%%%%%%%%%%%%%%%%%%%%%%%%%%%%%%%%%%%%%%%%%%%%%%%%%%%
\section{ Optional weights in  {\tt TAUOLA} Monte Carlo} 
%%%%%%%%%%%%%%%%%%%%%%%%%%%%%%%%%%%%%%%%%%%%%%%%%%%%%%%%%%%%%%%%%%%%%%%%%%%%%%%%%%%%%%%%%%%%%
Physics of $\tau$ lepton decays requires sophisticated strategies for the
confrontation of phenomenological models with experimental data. On one side 
high statistics experimental samples are collected, and the obtained precision is 
high, on the other hand, there is a significant cross-contamination between distinct
$\tau$ decay channels. Starting from  a certain precision  level all channels 
need to be analyzed simultaneously. Change of parametrization for one channel 
contributiong to the background may be important for the fit of another
 one. This situation leads to a complex configuatin where a multitude of parameters 
needs to be simultaneously confronted with a multitude of observables.
One has to keep in mind that the models used to obtain distributionbs in
 the fits may require refinements or even substantial rebuilds. 
The isospin relation to distribution measured in $e^+e^-$ collistions at low 
energies offer additional control, but may be the source of even more complex 
systematic error considerations. Effects of QED bremsstrahlung need to be 
taken ito account as well. It may affect selection criteria and background 
contaminations in quite complex and unexpected ways. 

From the statistical point of view it is best to resolve such system in one 
automated step using the method 
such as \cite{tmva,Hocker:2007ht} for example. 
This can be of course very dangerous from 
the point of view of systematic error control. We will return to this point 
later. At this moment let us discuss necessary adaptations for {\tt TAUOLA} 
for such methods to be used. In fact the principle of the changes is quite 
simple. The method is in use since long time and in different situations, 
see eg.~\cite{Jacholkowska:1999ei}.
 It is enough, to calculate, for each generated event (for each
present in it decay of $\tau^+$ and $\tau^-$ separately) alternative weights; the ratios
of matrix element squared obtained with new currents  provided by the user,
and the one actually used in generation are then  the vector of weights.
 If the currents 
can be represented as a linear combination of elementary ones, then 
the weight for arbitary values 
of coefficients is a bilinear form in the relative coefficients\footnote{We have 
applied
such method in one of our other programs \cite{Jacholkowska:1999ei} for the studies of anomalous couplings of $Z$ and LEP data.
}. In more general cases,
linearization is necessary, and the whole procedure needs to be repeated with 
the  currents obtained at one step used for generation at the next one. 
In this way, effects of the new currents will migrate to spin correlations 
between  $\tau^+$ and $\tau^-$ as well. 

\begin{figure}
\begin{center}
\setlength{\unitlength}{0.5 mm}
\begin{picture}(35,80)
%%%\put( 0,0){\framebox( 60,50){ }}
\put( -65,-45){\makebox(0,0)[lb]{\epsfig{file=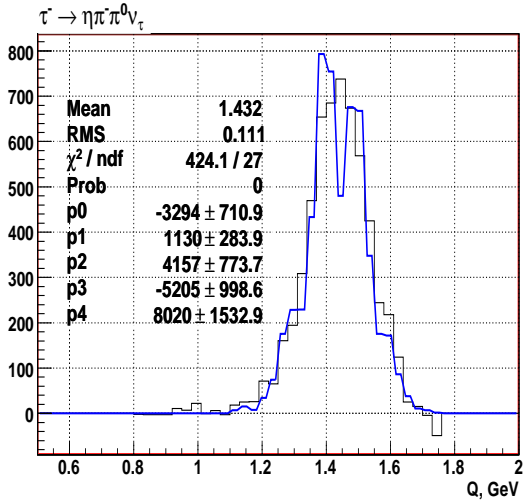,width=80mm,height=70mm}}}
\end{picture}
\end{center}
\vskip 1.5 cm
\caption{\small \it  Preliminary result from the study of 
$\tau^- \to \eta \pi^-\pi^0 \nu$ decay using external weight. Experimental 
data and theoretical predictions obtained at the intermediate step of the work 
are shown. The distribution of the invariant mass of all final state hadrons 
is given. This plot is available thanks to courtesy of Vladimir Cherepanov.
 } \label{vladimir}
\end{figure}

The first example of the use of the method, is demonstrated in fig.~\ref{vladimir}.
At this step of the work no detector effects were included. Experimental data 
were provided as a one-dimensional histogram. Simulation of $\tau$ production process was not performed. One can see, that the theoretical model provides double
peak which is absent in the data. This is because of an interplay between 
intermediate 
state resonance parametrization and phase space jacobian.
 The same parametrization was used for 3-scalar and 2-scalar systems. 
This points to the necessity of further investigations of unitarity constraints
 on the running width or better control of the data. 

Options how to handle, in experimental software, information 
for weight calculation is under study now.  
In principle, information stored in the event record should be enough. 
This may not be a perfect 
match in all cases. For example, effects of the boost from laboratory frame to
rest-frame of $\tau$ lepton and back, may result in rounding errors. 
In some cases the 
kinematical configuration may be changed because of QED bremsstrahlung 
generated by {\tt PHOTOS}. The order of final state particles stored in the event 
record may not be the same as  used internally in {\tt TAUOLA} at 
the event construction step. This of course can be cross checked on the basis 
of inspection  of a smaller sample and without detector simulation part being 
used. Such tests are nonetheless necessary, but may need less time than would 
be the case if collaboration software would be rebuild. From that point of 
view, the easiest would be to calculate optional weights from 
information already available
in experiments production tapes, and only for those events where bremsstrahlung 
effects are not generated at all. This can be correted at the next step of form-factor
iteration. Related slow down of the convergence should not be large.
Possibly two steps will be enough, the second one for the verification.
It will be necessary for control of spin correlations between the two decaying 
$\tau$-s as well. It is technically much simpler not to include  changes into 
spin correlations. In such a case the vector of weights would depend on kinematical 
configurations of $\tau^+$, $\tau^-$ and of the production process 
simultaneously.
The optimal choice require further interaction with collaborations.
Technical side of the activity will depend on the way the user want to calculate
the hadronic current. It can be easily calculated as external generalized function
communicating with {\tt TAUOLA} at the level of linker or through the 
{\tt fifo} pipes. They can be provided as {\tt FORTRAN} or eg. {\tt C++} code
and have well defined and easy to control input/output for modifications
introduced into {\tt TAUOLA}. 

Int the past, and for a good reason,
significant effort was devoted to strategies where the methotds as indicated 
here can be avoided. For example in \cite{Kuhn:1992nz} a system of projections 
was proposed for the 3-scalar $\tau$ decays. If such a method can be used
it is possible to obtain from the data just one of the contribution to the 
current, which is then parametrized by a single scalar function of two 
invariants only. This is of course on the expense of statistical strength on the 
measurement and abstraction of experimental sytematic error. Such methods, 
easy to apply for final states of two or three scalars, are less practical for 
higher multiplicities. Nonetheless they can be of great help to control 
the analysis and to verify which aspects of the models require improvements. 

%%%%%%%%%%%%%%%%%%%%%%%%%%%%%%%%%%%%%%%%%%%%%%%%%%%%%%%%%%%%%%%%%%%%%%%%%%%%%%%%%%%%%%%%%%%%%
\section{{\tt PHOTOS} and NLO effects in $W$ and $\gamma^* \to \pi^+\pi^-$ decays}
\def\CCol{{\tt SANC}}
%%%%%%%%%%%%%%%%%%%%%%%%%%%%%%%%%%%%%%%%%%%%%%%%%%%%%%%%%%%%%%%%%%%%%%%%%%%%%%%%%%%%%%%%%%%%%
The changes introduced over the last two years 
into  {\tt PHOTOS} Monte Carlo program itself were rather small. 
On the other hand, the complexity of the work on its theoretical foundation
was large and 
 matches neither size nor the  main purpose 
of the present talk (some of the related topics are revieved 
in~\cite{Was:2008zz}). 
Previous tests  of two body decays of the $Z$  
into a pair of charged leptons \cite{Golonka:2006tw}
and scalar $B$  into a pair of scalars  \cite{Nanava:2006vv} was recently 
supplemented  \cite{Photos_tests} with the study of $W^\pm \to l^\pm \nu \gamma$. 
The study of the process for $\gamma^* \to \pi^+\pi^-$ is on-going \cite{Xu}.
In all of these cases universal kernel of {\tt PHOTOS} was replaced with the 
one matching exact first order matrix element. In this way terms for NLO/NLL 
level are implemented. As usual, algorithm covers the full multiphoton 
phase-space and is exact in the infrared region of the phase space. 
This is rather unusual for the NLL compatible algorithms. One should not forget 
that {\tt PHOTOS} generates weight one events and does not require any 
factorization scales. There is full overlap of the phase space where hard 
matrix element is used with the one for iterative photon emissions.
All interference effects are implemented with the help of internal weights.

The results of all tests of {\tt PHOTOS} with a NLO kernel are
 at sub-permille level. No differences with 
benchmarks can be observed, even for samples of $10^9$ events. For the 
comparisons when simpler physics assumptions are used,
differences were sub-permille too, if compared to total rates 
(or were matching 
precision of programs used for tests). 

This is very encouraging, and points to the possible extension of the 
approach outside of the QED (scalar QED) only. In particular to the domain of 
QCD or if phenomenological lagrangians for interactions of photons need
 to be applied. For that work  to be completed spin amplitudes need 
to be further studied. Let us point here to the paper \cite{vanHameren:2008dy}.
 
The discussed here refinements
only indirectly affect practical side of simulations for $\tau$ physics. 
Changes  in the kernels necessary for NLO  may remain as 
options  for tests only. They are available from the {\tt PHOTOS} web page
\cite{Photos_tests} but are not recomended for wider use. An example of 
such new tests is
given in fig.~\ref{massquare}b.

%%%%%%%%%%%%%%%%%%%%%%%%%%%%%%%%%%%%%%%%%%%%%%%%%%%%%%%%%%%%%%%%%%%%%%%%%%%%%%%%%%%%%%%%%%%%%
\section{ { \tt MC-TESTER} and C++ {\tt TAUOLA universal interface}}
%%%%%%%%%%%%%%%%%%%%%%%%%%%%%%%%%%%%%%%%%%%%%%%%%%%%%%%%%%%%%%%%%%%%%%%%%%%%%%%%%%%%%%%%%%%%%

In the development of packages such as {\tt TAUOLA} or {\tt PHOTOS}, question 
of tests and appriopriate relations to users applications are essential for 
their 
usefulness. In fact, users applications may be much larger in size and 
human effort than the programs discussed here. 
Good example of such `user applications' are complete environments to fit 
experimental data and control detector uncertainties at the same time. 
As a consequence
our programs need to work well with {\tt FORTRAN} where {\tt HEPEVT} event record 
is used and {\tt C++} where {\tt HepMC} \cite{Dobbs:2001ck} is used instead.  For the sake of 
automatization of tests
{\tt MC-TESTER} was prepared~\cite{Golonka:2002rz}.
Recently, new functionalities
were introduced into the package \cite{Davidson:2008ma}. In particular, it works now with the
{\tt HepMC} event record, the  standard of {\tt C++} programs.
The complete set-up  for benchmarking the interfaces, such as interface
 between $\tau$-lepton
production and decay, including QED bremsstrahlung effects is prepared.
The  example is chosen to illustrate the new options introduced into the 
program and  novel ways of its use.
From the technical perspective, \cite{Davidson:2008ma}
documents program updates and  supplements previous documentation~\cite{Golonka:2002rz}.

As in the past,
test consists of two steps. Distinct Monte Carlo
programs are run separately; events  are searched for the decays of the 
chosen object, and  information is stored
 by {\tt MC-TESTER}. 
Then, at the analysis step,  information from a pair of such runs may
 be compared and represented in the form of tables and plots. 
\begin{figure}[!ht]
\setlength{\unitlength}{0.1mm}
\begin{picture}(800,1250)
%\put( 375,650){\makebox(0,0)[b]{\large }}
%\put( 375,1350){\makebox(0,0)[b]{\large }}
\put(5,-25){\makebox(0,0)[lb]{\epsfig{file=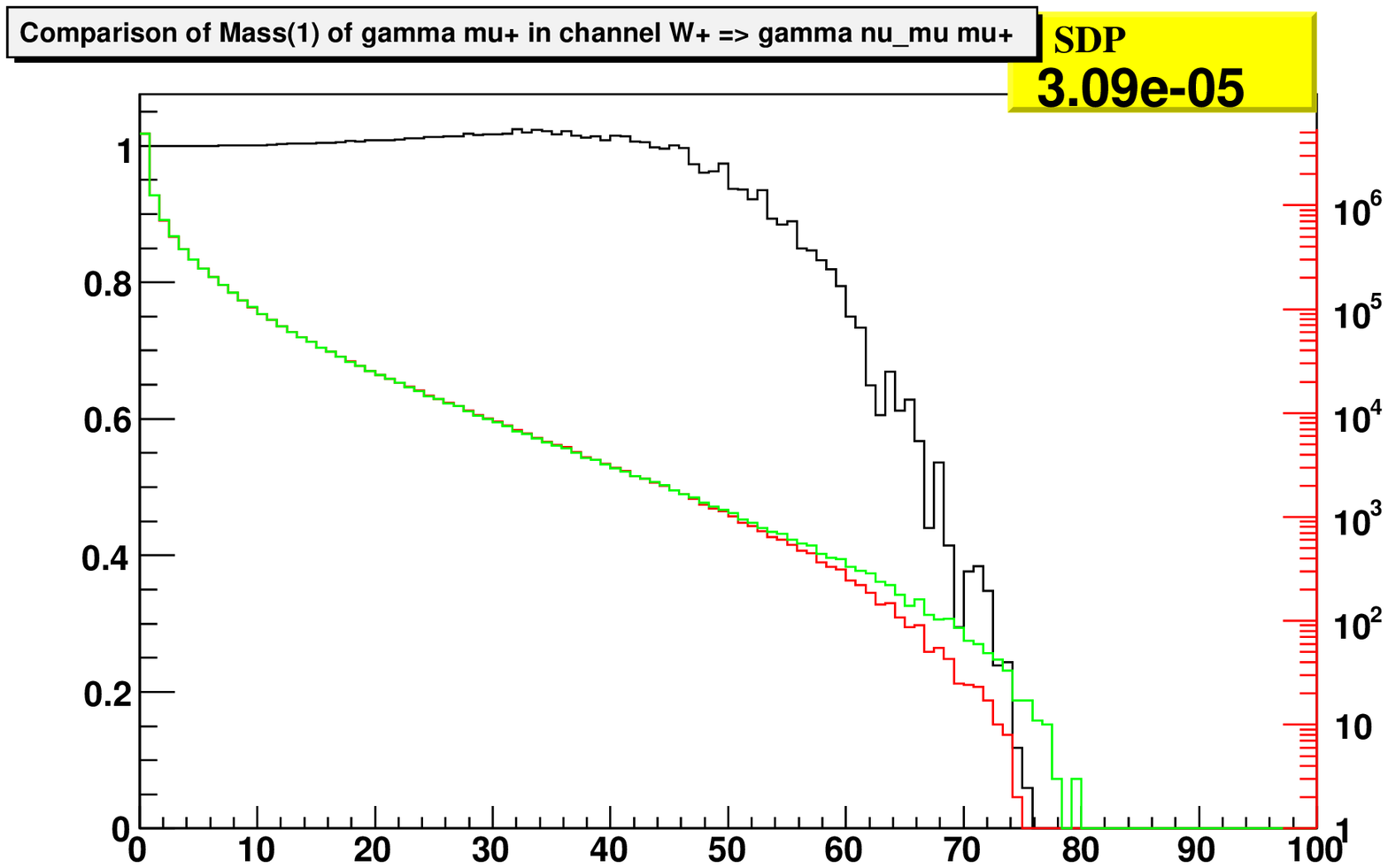,width=75mm,height=60mm}}}
\put( 55,-45){\makebox(0,0)[lb]{b}}
\put(5,650){\makebox(0,0)[lb]{\epsfig{file=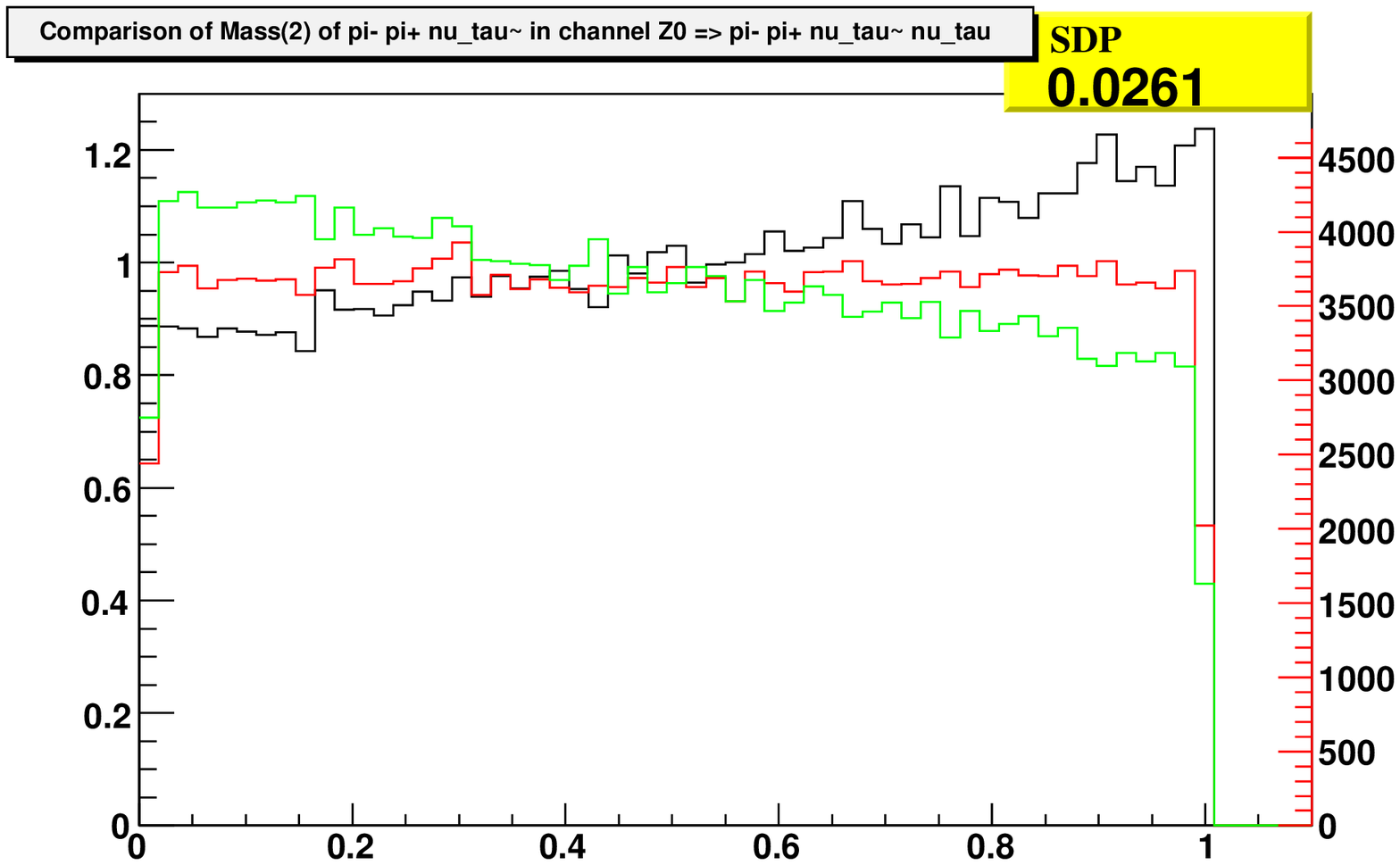,width=75mm,height=60mm}}}
\put( 55,650){\makebox(0,0)[lb]{a}}
\end{picture}
\caption 
{\it Examples of interesting benchmarks: In the decay
$Z \to \tau^+ \tau^- \to \pi^+ \pi^- \nu \bar \nu$, the $\pi^-$ energy spectrum in the Z rest-frame
has identical shape as the distribution of  the invariant mass squared 
 of $\pi^+ \pi^- \bar \nu$, see plot a. 
This linear distribution can be used to measure decaying $\tau^-$ polarization. 
In plot b, an example of the {\tt PHOTOS} test is given.
Program version with full matrix element for
$W^+ \to \mu^+ \nu_\mu \gamma$ channel is compared with the standard one. 
The  distribution of 
the $\mu^+ \gamma$ pair mass is shown.}
\label{massquare}
\end{figure}

One may have the impression that the modifications introduced into the  
release  are minor and consist of simple improvements in 
the graphical representation of the output and purely technical reorganization 
thanks to the use of {\tt C++} lists. 

To some degree this is true, but on the other hand the changes were introduced 
because of pressure from applications. Let us show 
how program modifications can be used for non-trivial practical applications.

It is quite common that information stored in the event record is too large. 
For example individual soft photons which remain undetectable are present. Not only
they do not influence  the detector response at all, but they exhibit technical 
aspects of eg. infrared regulators of QED bremsstrahlung. In response, 
{\tt MC-TESTER}
should ignore  those photons while analyzing decays 
(or group together with other particles). Otherwise, comparisons 
of different Monte Carlo programs would be dominated by the technical aspect of 
the implementation of the infrared regulator; {\tt MC-TESTER} operation need to be 
adopted for this, see eg. \cite{Golonka:2005pn,Golonka:2006tw}. 

Another example where the event tree may need to be simplified for validation is 
if spin correlations are appropriately introduced into various production 
processes.  Let us use as an example\footnote{Spin correlations  in 
  decays of $W,H,H^\pm$ into $\tau$ lepton(s)  are nearly identical.}
 $ pp \to Z/\gamma* + X$, 
$Z \to \tau^+ \tau^-$. It is convenient to start the test by restricting 
$\tau$ decays to the simplest decay mode, that is $\tau^\pm \to \pi^\pm \nu$,
and look at distributions in cascade decay $Z \to \pi^+ \pi^- \nu \bar \nu$.
In this case the effects of spin correlations are largest. 
The distribution of the $\pi^-$ energy spectrum (in the $Z$ rest-frame),
manifests the $\tau$ polarization through its slope (see 
fig  \ref{massquare}a). 
Fortunately, this frame dependent spectrum is equal to 
the  distribution of the square of invariant mass of $\pi^- \pi^+ \bar \nu$.
This distribution can be obtained in {\tt MC-TESTER} thanks to the new options.

Final state activities will lead to final states of $Z$ decays
where the $\tau$-pair is accompanied by bremsstrahlung photons or soft hadrons.
Such distribution are usually strongly peaked. In fig.~\ref{massquare}b we 
provide an example of such a distribution but for the comparison of 
{\tt PHOTOS}  with exact marix element for the
decay 
$W^+ \to \mu^+ \nu \gamma$. 

With the above brief examples 
we have demonstrated that   {\tt MC-TESTER} may be useful for tests of 
 libraries of particles decays, as well as for tests of their interfaces.
The work on {\tt HepMC } based  {\tt TAUOLA  universal interface} is 
on-going. A preliminary version is already available \cite{tauolaC++}, but from physics point of view it is still less complete than the {\tt FORTRAN} version~\cite{Pierzchala:2001gc}.

The updated version of {\tt MC-TESTER} was found \cite{Golonka:2005pn,Golonka:2006tw}
 to handle well
cases where physically spurious information (eg. on soft photons)
need to be ignored.

%%%%%%%%%%%%%%%%%%%%%%%%%%%%%%%%%%%%%%%%%%%%%%%%%%%%%%%%%%%%%%%%%%%%%%%%%%%%
%%%%%%%%%%%%%%%%%%%%%%%%%%%%%%%%%%%%%%%%%%%%%%%%%%%%%%%%%%%%%%%%%%%%%%%%%%%%
\section{Summary and future possibilities}

The status of the computer programs for the decay of $\tau$ leptons
{\tt TAUOLA}
 and
associated projects {\tt TAUOLA universal interface } and {\tt MC-TESTER}  
was reviewed.
The high-precision version of  {\tt PHOTOS} for radiative corrections was 
presented too.
 
New results for  {\tt PHOTOS} and  decay of $W$  were mentioned.
  For this channel complete next-to-leading order
effects can now be  simulated. However, for  most of the applications these effects
are not necessary, leaving the standard modular version of {\tt PHOTOS} sufficient.
This exercise is important not only for photonic bremsstrahlung. It helps 
to understand better questions related with matching hard emission matrix 
elements with parton showers without necessity to introduce any boundaries
within phase space.
 The important result of the above work is that the path to include 
electromagnetic form-factors of the particles participating in decay 
 is now open
for future fits to the data. These form-factor effects may be significantly larger 
and physically more justified than  complete next-to-leading order
effects of scalar QED as in $B$ meson decays.

 The presentation of the {\tt TAUOLA} general-purpose interface
  was brief. The {\tt C++} version exist, but is still physically less refined
than the {\tt FORTRAN} version.  Examples for its use can be found on
the continuously modified web
 page \cite{tauolaC++}. 
The new version of {\tt MC-TESTER} \cite{Davidson:2008ma} is already
public now. It not only works with {\tt HepMC} \cite{Dobbs:2001ck} of {\tt C++} but enables
user defined tests.  

Distinct versions of the {\tt TAUOLA} library for $\tau$ lepton decay, and of
{\tt PHOTOS} for radiative corrections in decays, are now in use.
 The principles how to use  the distribution
package are presented in refs. \cite{Golonka:2003xt,Was:2004dg}, on the 
other hand, development of new and recovery of codes for old strategies of fits is becoming 
available. This includes strategy to use weighted events, and the  projection 
operators \cite{Kuhn:1992nz} for the three scalars final states.  
Let us hope that the present talk will provide some stimulation for the larger 
work in the near future.

\vskip 2 mm
\centerline{ \bf Acknowledgements}
\vskip 1 mm

Discussions   
with  members of the Belle and BaBar collaborations 
are also acknowledged. Exchange of e-mails and direct discussions 
with  S. Banerjee, S. Eidelman, H. Hayashii, K. Inami,  A. Korchin, J. H. K\"uhn   and 
O. Shekhovtsova was a valuable  input to present and future steps in 
project development.

%%%%%%%%%%%%%%%%%%%%%%%%%%%%%%%%%%%%%%%%%%%%%%%%%%%%%%%%%%%%%%%%%%%%%%%%%%%%
%%%%%%%%%%%%%%%%%%%%%%%%%%%%%%%%%%%%%%%%%%%%%%%%%%%%%%%%%%%%%%%%%%%%%%%%%%%%
%\bibliographystyle{utphys_spires}
%\bibliographystyle{plain}
%\bibliography{TAUOLA-F}

\begingroup\begin{thebibliography}{10}

\bibitem{Jadach:1990mz}
S.~Jadach, J.~H. Kuhn, and Z.~Was, {\em Comput. Phys. Commun.} {\bf 64} (1990)
275.
%%CITATION = CPHCB,64,275;%%.

\bibitem{Jezabek:1991qp}
M.~Jezabek, Z.~Was, S.~Jadach, and J.~H. Kuhn, {\em Comput. Phys. Commun.} {\bf
  70} (1992)
69.
%%CITATION = CPHCB,70,69;%%.

\bibitem{Jadach:1993hs}
S.~Jadach, Z.~Was, R.~Decker, and J.~H. Kuhn, {\em Comput. Phys. Commun.} {\bf
  76} (1993)
361--380.
%%CITATION = CPHCB,76,361;%%.

\bibitem{Golonka:2003xt}
P.~Golonka {\em et al.}, {\em Comput. Phys. Commun.} {\bf 174} (2006) 818--835,
\href{http://www.arXiv.org/abs/hep-ph/0312240}{{\tt hep-ph/0312240}}.
%%CITATION = HEP-PH/0312240;%%.

\bibitem{Barberio:1990ms}
E.~Barberio, B.~van Eijk, and Z.~Was, {\em Comput. Phys. Commun.} {\bf 66}
  (1991)
115.
%%CITATION = CPHCB,66,115;%%.

\bibitem{Barberio:1994qi}
E.~Barberio and Z.~Was, {\em Comput. Phys. Commun.} {\bf 79} (1994)
291--308.
%%CITATION = CPHCB,79,291;%%.

\bibitem{tmva}
A.~Hoecker, P.~Speckmayer, J.~Stelzer, F.~Tegenfeldt, H.~Voss, K.~Voss, {\em et
  al.}, http://tmva.sourceforge.net/.

\bibitem{Hocker:2007ht}
A.~Hocker {\em et al.}, {\em PoS} {\bf ACAT} (2007) 040,
\href{http://www.arXiv.org/abs/physics/0703039}{{\tt physics/0703039}}.
%%CITATION = PHYSICS/0703039;%%.

\bibitem{Jacholkowska:1999ei}
A.~Jacholkowska, J.~Kalinowski, and Z.~Was, {\em Comput. Phys. Commun.} {\bf
  124} (2000) 238--242,
\href{http://www.arXiv.org/abs/hep-ph/9905225}{{\tt hep-ph/9905225}}.
%%CITATION = HEP-PH/9905225;%%.

\bibitem{Kuhn:1992nz}
J.~H. Kuhn and E.~Mirkes, {\em Z. Phys.} {\bf C56} (1992)
661--672.
%%CITATION = ZEPYA,C56,661;%%.

\bibitem{Was:2008zz}
Z.~Was, {\em Acta Phys. Polon.} {\bf B39} (2008) 1761,
\href{http://www.arXiv.org/abs/0807.2775}{{\tt 0807.2775}}.
%%CITATION = 0807.2775;%%.

\bibitem{Golonka:2006tw}
P.~Golonka and Z.~Was, {\em Eur. Phys. J.} {\bf C50} (2007) 53--62,
\href{http://www.arXiv.org/abs/hep-ph/0604232}{{\tt hep-ph/0604232}}.
%%CITATION = HEP-PH/0604232;%%.

\bibitem{Nanava:2006vv}
G.~Nanava and Z.~Was, {\em Eur. Phys. J.} {\bf C51} (2007) 569--583,
\href{http://www.arXiv.org/abs/hep-ph/0607019}{{\tt hep-ph/0607019}}.
%%CITATION = HEP-PH/0607019;%%.

\bibitem{Photos_tests}
P.~Golonka, G.~Nanava, and Z.~Was, Tests of PHOTOS Hard Bremsstrahlung,
  http://mc-tester.web.cern.ch/MC-TESTER/PHOTOS-MCTESTER/.

\bibitem{Xu}
Z.~Was and Q.~Xu, In preparation.

\bibitem{vanHameren:2008dy}
A.~van Hameren and Z.~Was,
\href{http://www.arXiv.org/abs/0802.2182}{{\tt 0802.2182}}.
%%CITATION = 0802.2182;%%.

\bibitem{Dobbs:2001ck}
M.~Dobbs and J.~B. Hansen, {\em Comput. Phys. Commun.} {\bf 134} (2001) 41--46,
https://savannah.cern.ch/projects/hepmc/.
%%CITATION = CPHCB,134,41;%%.

\bibitem{Golonka:2002rz}
P.~Golonka, T.~Pierzchala, and Z.~Was, {\em Comput. Phys. Commun.} {\bf 157}
  (2004) 39--62,
\href{http://www.arXiv.org/abs/hep-ph/0210252}{{\tt hep-ph/0210252}}.
%%CITATION = HEP-PH/0210252;%%.

\bibitem{Davidson:2008ma}
N.~Davidson, P.~Golonka, T.~Przedzinski, and Z.~Was,
\href{http://www.arXiv.org/abs/0812.3215}{{\tt 0812.3215}}.
%%CITATION = 0812.3215;%%.

\bibitem{Golonka:2005pn}
P.~Golonka and Z.~Was, {\em Eur. Phys. J.} {\bf C45} (2006) 97--107,
\href{http://www.arXiv.org/abs/hep-ph/0506026}{{\tt hep-ph/0506026}}.
%%CITATION = HEP-PH/0506026;%%.

\bibitem{tauolaC++}
N.~Davidson, T.~Przedzinski, E.~Richter-Was, and Z.~Was, Presently the code is
  evolving quickly. Interested pilot users can download code from
  http://www.ph.unimelb.edu.au/ \~{}ndavidson/tauola/doxygen/index.html It is
  automatically updated daily from our source code repository. Version
  information can be found in a text file contained in the distribution.

\bibitem{Pierzchala:2001gc}
T.~Pierzcha\l{a}, E.~Richter-W\c{a}s, Z.~W\c{a}s, and M.~Worek, {\em Acta Phys.
  Polon.} {\bf B32} (2001) 1277--1296,
\href{http://arXiv.org/abs/hep-ph/0101311}{{\tt hep-ph/0101311}}.
%%CITATION = HEP-PH 0101311;%%.

\bibitem{Was:2004dg}
Z.~Was and P.~Golonka, {\em Nucl. Phys. Proc. Suppl.} {\bf 144} (2005) 88--94,
\href{http://www.arXiv.org/abs/hep-ph/0411377}{{\tt hep-ph/0411377}}.
%%CITATION = HEP-PH/0411377;%%.

\end{thebibliography}\endgroup
\providecommand{\href}[2]{#2}                                    %%%%%%%%%%%%%%%%%%
\end{document}